\documentclass[aps,prl,twocolumn,showpacs,groupedaddress]{revtex4}
\usepackage[dvips]{graphicx}
\usepackage{amssymb}
\begin{document}
%\preprint{}
\title{Andreev Probe of Persistent Current States in
Superconducting Quantum Circuits}
\author{V. T. Petrashov}
\author{K. G. Chua}
\author{K. M. Marshall}
\author{R. Sh. Shaikhaidarov}
\author{J. T. Nicholls}
\affiliation{Department of Physics, Royal Holloway, University of
London, Egham, Surrey TW20 0EX, UK}
\date{\today}
\begin{abstract}
Using the extraordinary sensitivity of Andreev interferometers to
the superconducting phase difference associated with currents, we
measure the persistent current quantum states in superconducting
loops interrupted by Josephson junctions. Straightforward
electrical resistance measurements of the interferometers give
continuous read-out of the states, allowing us to construct the
energy spectrum of the quantum circuit. The probe is estimated to
be more precise and faster than previous methods, and can measure
the local phase difference in a wide range of superconducting
circuits.
\end{abstract}

\pacs{03.67.Lx, 85.25.Cp , 85.25.Dq} \maketitle

Superconducting circuits consisting of loops interrupted by
Josephson junctions show persistent current states that are
promising for implementation in a quantum computer \cite{mooij99}.
Spectroscopy and coherent quantum dynamics of the circuits have
been successfully investigated by determining the
switching-to-voltage-state-probability of an attached
superconducting quantum interference device (SQUID)
\cite{chior03}; however, a single switching measurement is low
resolution and strongly disturbs both the circuit and the SQUID
itself. This revives the fundamental problem of fast high
resolution quantum measurements of the persistent current states.
The conceptual and technological advance reported here is based on
the fact that a persistent current in a quantum circuit is
associated with the gradient of the superconducting phase $\chi$
of the macroscopic wavefunction describing the circuit. The
problem of measuring the current reduces to a measurement of the
corresponding phase difference $\theta_q$ across the Josephson
junctions.

To measure $\theta_q$ with a minimum of disruption we use an
Andreev interferometer \cite{pet94,naz96}. Our Andreev
interferometers, shown in the scanning electron microscope images
in Figs.~1a and b, are crossed normal ($N$) silver conductors
$a$-$b$ and $c$-$d$, with contacts to a pair of superconducting
($S$) aluminium wires at the points $c$ and $d$. The $N/S$
interfaces play the role of mirrors reflecting electrons via an
unusual mechanism first described by Andreev \cite{andreev64}. In
Andreev reflection, an electron which is incident on the normal
side of the $N/S$ interface evolves into a hole, which retraces
the electron trajectory on the $N$-side, and a Cooper pair is
created on the $S$-side. There is a fundamental relationship
between the macroscopic phase of the superconductors and the
microscopic phase of the quasiparticles \cite{spiv82}: the hole
gains an extra phase equal to the macroscopic phase $\chi$, and
correspondingly the electron acquires an extra phase $-\chi$. This
leads to phase-periodic oscillations in the resistance $R_A$
between the points $a$ and $b$ of the interferometer. It should be
emphasized that the macroscopic phase is probed by quasiparticles
with energies much less than the superconducting gap, so there is
no ``quasiparticle poisoning'' of the superconductor.

\begin{figure}[b]
\label{f:01}
\begin{center}
\includegraphics[angle=0,width=6cm,keepaspectratio,clip]{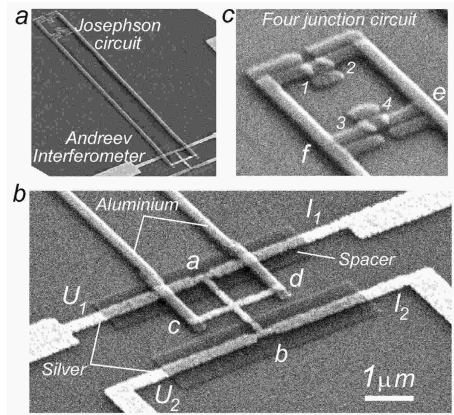}
\end{center}
\caption{Layout of Josephson circuit with attached Andreev
interferometer. (a) General view. (b) Andreev interferometer
consisting of crossed silver wires, connected to the aluminium
wires at $c$ and $d$. The resistance $R_A$ between $a$ and $b$ is
measured using current ($I_1$, $I_2$) and voltage probes ($U_1$,
$U_2$).  (c) The superconducting quantum loop is interrupted by
Josephson junctions at points 1, 2, 3, and 4. There is a
superconducting phase difference $\theta_q$ between $e$ and $f$.}
\end{figure}

We investigate a Josephson quantum circuit with an attached
Andreev interferometer, as shown in Figs.~1a and c.  To probe the
phase difference within the Josephson circuit, superconducting
wires were connected to the points $e$ and $f$, as shown in
Fig.~1c. The four-terminal resistance $R_A$ was measured using the
current ($I_1$, $I_2$) and voltage ($U_1$, $U_2$) probes shown in
Fig.~1b. The oscillating part of the resistance $\delta R_A$
depends on the superconducting phase difference $\phi$ between $c$
and $d$, which can be described by \cite{pet98}
\begin{equation}\label{e:1}
\delta R_A = - \gamma \cos \phi,
\end{equation}
where the amplitude $\gamma$ is independent of $\phi$. The phase
difference $\phi$ can be written \cite{Lik86} as $\phi = 2 \pi
\frac{\Phi_A}{\Phi_0} + \theta_q$, where $\theta_q$ is the phase
difference between the points $e$ and $f$ introduced by the lower
branch of the Josephson loop. Due to a magnetic field $B$ applied
perpendicular to the plane of the device, the total flux through
the interferometer area $S_A$ (enclosed by $c$-$d$-$e$-$f$) is
$\Phi_A = \Phi_{eA} - L_{A}I_{SA}$, where $\Phi_{eA}=S_A B$ is the
external flux through $S_A$ which has an inductance $L_A$.
$I_{SA}$ is the current circulating in the interferometer loop,
and $\Phi_0$ is the flux quantum $h/2e$.

Our Andreev probes were designed according to three
criteria: \\ %
\noindent {\bf I.} To exclude parasitic potential differences
between the $N/S$ interfaces, we fabricate interferometer
structures that are symmetric crosses.\\ %
\noindent {\bf II.} We ensure that the critical current induced in
the normal wires (and hence the current $I_{SA}$ circulating in
the interferometer loop) is zero. Thus we exclude both the direct
influence of $I_{SA}$ on the superconducting circuit, as well as
the back-action of the measuring current $I_m$. According to
experimental \cite{basel99,shaik00} and theoretical
\cite{volkov95,wilhelm98,yip98} studies the influence of the
current through $a$-$b$ on the superconductors connected at
$c$-$d$ vanishes when the critical current is zero.\\ %
\noindent {\bf III.} To suppress $I_{SA}$, but maintain the
sensitivity of the conductance to phase, the length $L_{cd}$ must
satisfy the condition  $\xi_N < L_{cd} < L_{\phi}$, where $\xi_N =
\sqrt{\hbar D/2 \pi k_B T}$ and $L_{\phi}=\sqrt{D \tau_\phi}$ are
the coherence length and the phase breaking length of the normal
metal, respectively; $D$ is the diffusion coefficient, and
$\tau_\phi$ is the normal metal phase breaking time. The critical
current is a thermodynamic property with contributions from
quasiparticles within $k_{B}T$ of the Fermi energy, and decays
within the coherence length. In contrast, the phase coherent
conductance is a kinetic property with contributions within the
Thouless energy $E_{Th}= h D/L_{cd}^2$, and survives up to the
order of $L_{\phi}$ \cite{pet98, volkov96, golubov97}. In this
limit the Josephson circuit phase is given by
\begin{equation}\label{e:3}
\theta_q = \phi - 2 \pi \frac{\Phi_{eA}}{\Phi_0},
\end{equation}
and {\em does not depend on measurement details}.

We have tested Andreev probes on three-junction~\cite{mooij99} and
four-junction Josephson circuits, and have found qualitatively
similar behaviour for both circuits. %
Four-junction circuits allow a symmetric connection to the
interferometer, which we believe minimizes the effect of
noise currents in the interferometer loop on the quantum states. %
The devices were fabricated using three-layer electron beam
lithography on silicon substrates covered with native oxide. %
The silver wires of the interferometer are 40 nm thick and 240~nm
wide, and the aluminium superconducting wires are 35~nm thick and
360~nm wide. The Josephson circuits are also aluminium,
interrupted by Al$_2$O$_3$ Josephson junctions (Figs.~1 a,c). The
spacer was a 30~nm thick Al$_2$O$_3$ film (Fig. 1b). Resistances
were measured using standard low frequency techniques at
temperatures between 0.02- 1.2~K.

\begin{figure}[ht]
\label{f:02}
\begin{center}
\includegraphics[angle=0,width=5.0cm,keepaspectratio,clip]{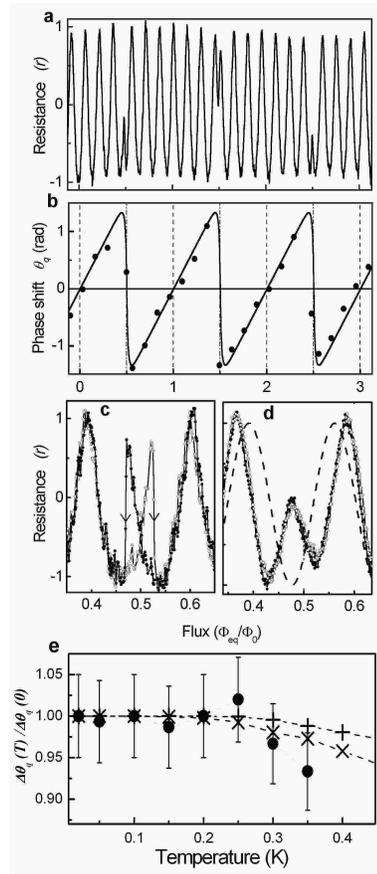}
\end{center}
\caption{(a) Normalized oscillating resistance, $r=\delta
R_A/\gamma$, measured at 20~mK. %
(b) The phase shift $\theta_q(\Phi_{eq})$ between $e$ and $f$, as
extracted from the oscillations in Fig.~2a; black dots are
experimental data and the solid line is calculated using
Eqs.~\ref{e:4} and \ref{e:5}. %
(c) Hysteresis in the resistance at the degeneracy point
$\Phi_{eq}=\Phi_0/2$ in an interferometer attached to a
``classical'' Josephson circuit, taken with increasing ($\circ$)
and decreasing ($\bullet$) magnetic field. %
(d) Detail of oscillations in Fig.~2a near the degeneracy point
$\Phi_{eq}=\Phi_0/2$, taken with increasing ($\circ$) and
decreasing ($\bullet$) magnetic field. The dashed line corresponds
to $\phi= 2 \pi \frac{\Phi_{eA}}{\Phi_0}$. %
(e) Experimental temperature dependence of the amplitude of
$\theta_q(\Phi_{eq})$ ($\bullet$), compared to calculations for
$\Delta =0.035E_J$ ($+$) and $\Delta=0.025 E_J$ ($\times$).}
\end{figure}

Figure 2a shows the normalized resistance $r=\delta R_A/\gamma$ of
an interferometer with $L_{cd} = 2$~$\mu$m, measured as a function
of normalized magnetic flux $\Phi_{eq}/\Phi_0$, where $\Phi_{eq}=B
S_q$ is the external flux through the Josephson junction loop area
$S_q = 2.45 \times 2.45~\mu$m$^2$. Using a test structure the
circulating current $I_{SA}$ in the interferometer loop was
measured to be zero, in agreement with a coherence length in
silver of $\xi_N \leq 1~\mu$m (estimated from $D \approx
100$~cm$^2$/s, which was obtained from resistivity measurements).
Interference oscillations are measured with a period close to
$\Phi_0$ through the interferometer area $S_A = 2.45 \times
18.4~\mu$m$^2$. The amplitude $\gamma$ depends on the resistance
$R_b$ of the $N/S$ interfaces, reaching values up to $0.12 R_A$
\cite{pet94}. In this particular device $R_A=5$~$\Omega$ and
$\gamma \approx 0.1~\Omega$. The measurement current was $I_m
=1-5~\mu$A, and the magnetic flux induced by this current was
negligible. Figure~2a shows there are abrupt phase shifts when the
flux $\Phi_{eq}$ corresponds to an odd number of half flux quanta,
$\Phi_{eq}^n=(2 n +1)\Phi_0/2$, where $n$ is an integer. The
dependence of the phase $\theta_q$ on magnetic flux is shown in
Fig.~2b; the sawtooth structure results from a build-up of
persistent current in the Josephson loop, followed by a transition
between states of different circulation.

\begin{figure}
\label{f:03}
\begin{center}
\includegraphics[angle=0,width=5.0cm,keepaspectratio,clip]{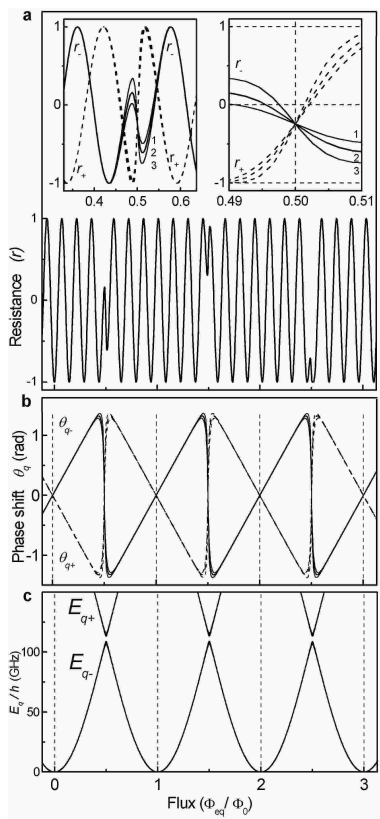}
\end{center}
\caption{Calculated fits to the experimental data of Figs.~2a, b
and d using Eqs.~\ref{e:4} and \ref{e:5}. (a) Calculated
oscillations of the normalized resistance $r(\Phi_{eq})$ in the
ground energy state as a function of the normalized external flux
$\Phi_{eq}/\Phi_0$. The insets show detail of oscillations in the
ground (solid lines) and excited states (dashed lines) for
$\Delta= 0.04 E_J$, $\Delta= 0.03 E_J$, and $\Delta= 0.02 E_J$
(curves 1, 2, and 3). (b) Calculated phase shifts
$\theta_{-}(\Phi_{eq})$ and $\theta_{+}(\Phi_{eq})$ in the ground
and excited states. (c) The energy spectrum $E_q$.}
\end{figure}

The shape of the transition at $\Phi_{eq}^n$ depends on parameters
of the Josephson circuit. Figure~2c shows transitions in a circuit
with inductance $L_q \approx 0.5$~nH, and a high critical current,
$I_{cq} \approx 1~\mu$A. There is hysteresis associated with the
transitions from clockwise to anticlockwise persistent current
states; this is the classical regime where the Josephson energy
(and the potential barrier between the persistent current quantum
states) is so high that there is no quantum tunnelling at
$\Phi_{eq}^n$. Figure~2d shows a close-up of the oscillations in
Fig.~2a, measured in a Josephson circuit with a lower critical
current, $I_{cq} \approx 0.1~\mu$A; there is a
smooth switch from one state to another, with no evidence of hysteresis. %
Also shown in Fig.~2d is a dashed line that corresponds  %
to $\phi =2 \pi \frac{\Phi_{eA}}{\Phi_0}$, which crosses the
measured curve $r(\Phi_{eq})$ at $\Phi_{eq}=\Phi_{0}/2$, the flux
at which $\theta_q=0$.

We have measured the influence of the measuring current $I_m$ on
the phase $\theta_q$. To within the accuracy of our measurements
(less than 5\%), the amplitude of $\theta_q$ is unaffected by
currents up to $I_m = 5~\mu$A (corresponding to 25~$\mu$V across
$a$-$b$). High $I_m$ currents may also induce thermal effects;
however, as shown in Fig.~2e the amplitude of $\theta_{q}$ is
constant over a wide range of temperatures.

Our phase measurements allow us to investigate the energy spectrum
of the Josephson circuit. From the equation
$\theta_q=\sin^{-1}\left( \frac{I_{Sq}}{I_{cq}} \right)$, the
phase difference $\theta_q$ across the Josephson junction is
related to the persistent current $I_{Sq}$ in the Josephson loop;
$I_{Sq}$ is itself related to the energy $E_q$ of the Josephson
loop through the derivative $I_{Sq}=\frac{\partial E_q}{\partial
\Phi_{eq}}$. Therefore, the equation
\begin{equation}\label{e:4}
\theta_q = \sin^{-1} \left(\frac{1}{I_{cq}}\frac{\partial
E_q}{\partial \Phi_{eq}}\right)
\end{equation}
shows that $\theta_q(\Phi_{eq})$ measurements allow the
determination of the energy spectrum $E_q(\Phi_{eq})$. To
demonstrate the technique, we use a generic form for the spectrum
\begin{widetext}
\begin{equation}\label{e:5}
E_q = \frac{\epsilon_q(\Phi_{eq})+\epsilon_q(\Phi_{eq}-\Phi_0)}{2}
\pm
\sqrt{\left(\frac{\epsilon_q(\Phi_{eq})-\epsilon_q(\Phi_{eq}-\Phi_0)}{2}\right)^2+
\Delta^2}
\end{equation}
\end{widetext}
where $2\Delta$ is the energy gap at $\Phi_{eq}=\Phi_0/2$ between
the excited ($E_{q+}$) and ground ($E_{q-}$) states. For energies
far away from the degeneracy point $\Phi_{eq}=\Phi_0/2$, we take
the junction charging energy $E_c = e^2/2C$ ($C$ is the junction
capacitance) to be much less than the Josephson energy $E_J$ (test
structures show that $E_J/E_C \approx 10$, with $E_J/h \approx
100$~GHz and $E_C/h = 12$~GHz). Then $\epsilon_q$ can be modelled
with the two-junction energy $\epsilon_q = E_J \left[1-\cos\left(\
\pi \frac{\Phi_{eq}}{\Phi_0} \right) \right]$, where $E_J = 2
I_{cq} \Phi_0/ 2 \pi$. To fit our measurements (Fig. 2) of the
ground state $E_{q-}$ there is only one free parameter $\Delta$;
we find that $\Delta = (0.03 \pm 0.005)E_J$ produces the best fit
to $r(\Phi_{eq})$ and $\theta_q$ over a wide range of flux, as
well as generating the energy spectrum $E_q$ in Fig.~3c. %
The model can also describe the temperature dependence of the
amplitude of $\theta_q$ shown in Fig.~2e; the reduction of
$I_{Sq}$ due to thermal fluctuations is given by $I_{Sq}=
I_{Sq}(0) \tanh [(E_{q+}-E_{q-})/2 k_B T]$, and the resulting
calculated amplitude of $\theta_q(T)$ is shown as dashed lines in
Fig.~2e for $\Delta =0.035E_J$ ($+$) and $\Delta=0.025 E_J$
($\times$).

In anticipation of measurements of the excited states, we use the
spectrum $E_q$ to calculate, see insets of Fig.~3a, the resistance
of the ground state $r_{-}(\Phi_{eq})$ and excited state
$r_{+}(\Phi_{eq})$.  When the circuit is irradiated with frequency
$\omega_0= (E_{q+}-E_{q-})/\hbar$, the measured voltage is
expected to oscillate at the Rabi frequency with an amplitude
$\Delta V_A =I_m (r_{+}(\Phi_{eq})-r_{-}(\Phi_{eq}))$.

The probe has an operating range from DC to an upper frequency,
$f_0$, which is limited by the quasiparticle's finite
time-of-flight between the N/S interfaces. For our probe $f_0
\approx  D/ L_{cd}^2 \approx 10$~GHz. The wide frequency response
allows measurements in both the continuous ``Rabi spectroscopy''
regime \cite{ilichev03} and the pulse regime \cite{chior03,lup04}.
Note, the Andreev probe measures local phase differences, enabling
the direct determination of quantum entanglement between different
elements of complicated Josephson circuits, which could be
unattainable with previous methods \cite{ilichev03,lup04}. An
increase in the operation speed by orders of magnitude can be
achieved using ballistic Andreev interferometers made using a high
mobility two-dimensional electron gas (2DEG), which will also
allow gate-controlled Andreev probes. Additionally, probes can be
fabricated to be impedance matched to standard 50~$\Omega$ or
75~$\Omega$ high frequency setups.

From our measurements we estimate the efficiency of the Andreev
probe compared to other methods. The signal-to-noise ratio ($SNR$)
for continuous measurements over a frequency range $\delta f$ is
$SNR=\Delta V_A/\sqrt{S_V \delta f}$, where $S_V$ is the spectral
density of the voltage noise. %
With $R_A=50~\Omega$, %
$\Delta R_A= r_{+}(\Phi_{eq})-r_{-}(\Phi_{eq}) = 1~\Omega$, %
$I_m=5~\mu$A, $\delta f \sim 2$~kHz, and with the noise
temperature of the cold amplifier $T_N \approx ~1$~K used in
\cite{ilichev03} we obtain %
$SNR=\Delta R_A I_m/\sqrt{4 k_B T_N \delta f} \approx 10^3$ for
the thermal noise, which is two orders of magnitude larger than
previously reported \cite{ilichev03}.

For the pulse technique an important parameter is the
discrimination time $\tau_m$, which is the time required to obtain
enough information to infer the quantum state. For reflection
measurements, the ``single shot'' measurement time is calculated
to be $\tau_m = S_V/(\Delta V_R)^2$, where $\Delta V_R= {{\partial
V_R} \over {\partial \Gamma}} {{\partial \Gamma} \over {\partial
R_A}} {{\partial R_A} \over {\partial \theta_q}} \Delta \theta_q$
is the reflected signal, $V_R= \Gamma V_A$, where $V_A=I_m R_A$
and $\Gamma = {{R_A - Z_0} \over {R_A+ Z_0}}$. %
Substituting the cold amplifier noise temperature $T_N
\approx 20$~K used in \cite{lup04}, %
$\Delta R_A = {{\partial R_A} \over {\partial \theta_q}}
(\theta_{q^+} - \theta_{q^-}) \approx 1~\Omega$,
$R_A=Z_0=50$~$\Omega$, $I_m=5$~$\mu$A, we estimate $\tau_m \approx
8.8 \times 10^{-7}$~s for the thermal noise - this is more than an
order of magnitude shorter than reported in \cite{lup04}. $\tau_m$
can be further improved using lower noise cryogenic amplifiers.
Measurements of the excited states will reveal the actual
decoherence mechanisms. In the mesoscopic interferometer the
thermal noise current can be minimized by reducing the number of
conducting channels in the length $L_{cd}$.

In summary, simple resistance measurements of an Andreev
interferometer provide direct read-out of the local
superconducting phase difference in quantum circuits; within the
accuracy of existing theory there is negligible back-action on the
quantum circuit. From the phase $\theta_q$, the energy spectrum
$E_q$ can be constructed. The probe is expected to be more precise
and faster than previous methods \cite{ilichev03,lup04}, and can
measure the local phase difference in a wide range of
superconducting circuits. The 2DEG-based Andreev probe can be made
gate-controlled. Our probe will allow us to address fundamental
aspects of quantum measurements. As the operator of the average
phase commutes with the two-state Hamiltonian,  measuring the
average phase may enable realization of ``quantum non-demolition''
(QND) measurements~\cite{braginsky96}, possessing important
features such as an accuracy that exceeds quantum limits
\cite{averin02}.

We thank Yu.~V. Nazarov, D.~V. Averin, P.~Delsing, M.~Lea, A.~F.
Volkov, D.~Esteve, D.~Vion, H.~Pothier, A. Zagoskin, and A.~M. van
den Brink for valuable discussions. This work was supported by the
Engineering and Physical Sciences Research Council (UK).

% \bibliography{vtp1}
% \end{document}

% \bibliography{vtp1}
\end{document}